\documentclass[]{aa} 

\usepackage{graphicx}
\usepackage{txfonts}
\begin{document}

   \title{The star in RNO 54 - a post-FUor with short faint jet?}

   \author{T.Yu. Magakian
          \inst{1}
          \and
          T.A. Movsessian
          \inst{1}
          \and
          H.R. Andreasyan
          \inst{1}
          \and
          A.V. Moiseev
          \inst{2}
          \and R.I. Uklein\inst{2}}

   \institute{Byurakan Astrophysical Observatory, 0213 Aragatsotn prov., Armenia\\
              \email{tigmag@sci.am}
         \and
             Special Astrophysical Observatory, N.Arhyz, Karachaevo-Cherkesia, 369167 Russia\\
               \email{moisav@sao.ru}
       }
   \date{Received ; accepted }

 
  \abstract
   {}
   {The aim of the present study was the spectral analysis of an unusual pre-main-sequence star in the cometary nebula RNO 54, which was suspected by several researchers as a FUori-like object. }
   {We performed long-slit spectroscopy of the star on the 6-m
telescope with the SCORPIO-2 multi-mode focal reducer.}
   {We discover a short ($\sim4$\arcsec\ or $\sim6000$  AU) and faint emission shock-excited jet from this star, probably oriented toward the long axis of the nebular ellipse. The spectral type of the star is estimated as G0-2 II; the split of absorption \ion{Li}{i} line which is a typical sign indicating the FUori-like spectrum, is confirmed. The analysis of the available data shows virtual absence of the photometric variability, for at least the last 20 years. The lower limit of the bolometric luminosity of the star is estimated as 300 $L_\sun$. Our study supports the classification of  RNO 54 star as a  FUor-like object in the long-after-outburst stage. } 
   {}


   \maketitle
%

\section{Introduction}

The  nebulous object RNO 54 was for the first time described in the list of \cite{RNO}. It appears as a bright star on the edge of an elliptic loop-like reflection nebula. In the same catalogue the star was classified as F5~II, with H${\alpha}$ emission. The further attention to this object was drawn by \cite{Goodrich}, who compared its morphology with ring-shaped nebulae near FUors. A new estimate of the spectral type of the central star, made by Herbig  on the base of the spectra in yellow-red range \citep[not shown in the work of][]{Goodrich}, was early-G Ib-II. 

This object is located on the edge of dark cloud Dobashi 4549 (TGU 1314). Its central star was also identified as  an infrared source IRAS 05393+2235 while in 2MASS catalog it is designated as 2MASS J05422123+2236471. Since RNO 54 name refers in fact to the nebulous object as whole, below we will use for its central star the 2MASS J05422123+2236471 designation.

In fact, the only previous spectrum of this object, published so far, is shown in the work of \cite{Torres}. In this work the split of \ion{Li}{I} $\lambda$ 6707 \AA\  line was pointed out as a further argument for the probable FUor-like nature of this star. Also its spectral energy distribution (SED) was presented. It demonstrates significant far-IR excess. The searches of extended shock-excited emission (i.e. Herbig-Haro knots or flows) in the vicinities of RNO~54, performed by the authors with the 2.6-m telescope of Byurakan observatory, were unsuccessful. The same negative result was recently obtained by \cite{Lopez}.

In this work we describe our new spectral observations of 2MASS J05422123+223647.


\section{Observations}

We performed the long-slit spectroscopy of 2MASS J05422123+223647 with the SCORPIO-2 multi-mode focal reducer \citep{scorpio2} at the Special Astrophysical Observatory of Russian Academy of Sciences (SAO RAS) 6-m telescope. We obtained two spectra on the nights Nov 25/26 and Dec 30/31 in 2021. The slit was oriented along the long axis of the nebula, with position angle PA = 10\degr. The total exposure time were 1200 s and 720 s in non-photometric atmospheric conditions, but with a relatively good seeing value $\sim1.1$\arcsec and 1.3$\arcsec$.  The spectrograph's slit with a width 1\arcsec\ and a length 6\arcmin\ gave the spectral resolution ($FWHM$)\textbf{ of }about 2.5\AA\ in the spectral range 5580-7750 \AA. The detector E2V CCD 261-84 with a format $4K \times2K$ was operated in $1\times2$ binning mode. It provides an angular scale along the axis $0.4$\arcsec per px with a mean dispersion $0.53$\AA\ per px.
The spectral data reduction was performed in a standard way, using the IDL-based software as described in our previous papers \citep[see, for example][]{Egorov2018}. It was not possible to perform the photometric calibration of the first spectrogram (Nov 25) due to bad weather conditions, but it was used for the radial velocity measurements.


\section{Results}

\begin{figure}[h!]
  \centering
  \includegraphics[width=0.5\textwidth]{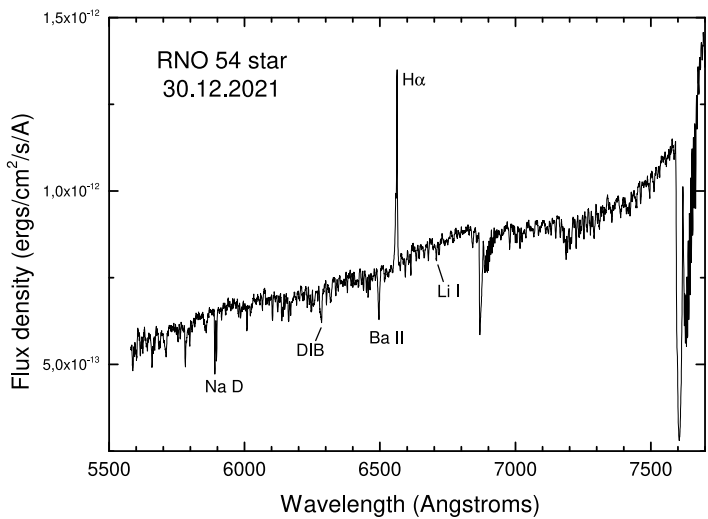}
   \caption{The spectrum of 2MASS J05422123+223647. }
   \label{star}
\end{figure}

In Fig.\ref{star} we present the spectrum of 2MASS J05422123+223647, obtained on 2021 Dec 30. It is rich with narrow absorption lines and has emission only in H$\alpha$ line. Comparison with the spectral library of  \citet{jacoby} shows that the spectral type of this star corresponds to the slightly reddened early G supergiant, in full accordance with the earlier estimate made by Herbig (see Sect.1). It should be noted that besides of stellar lines, the   diffuse interstellar band $\lambda$ 6284 \AA, well detached from the atmospheric absorption  $\lambda$ 6279 \AA\ is prominent in the spectrum of the 2MASS J05422123+223647. 

\begin{figure*}[h!]
  \centering
  \includegraphics[width=0.9\textwidth]{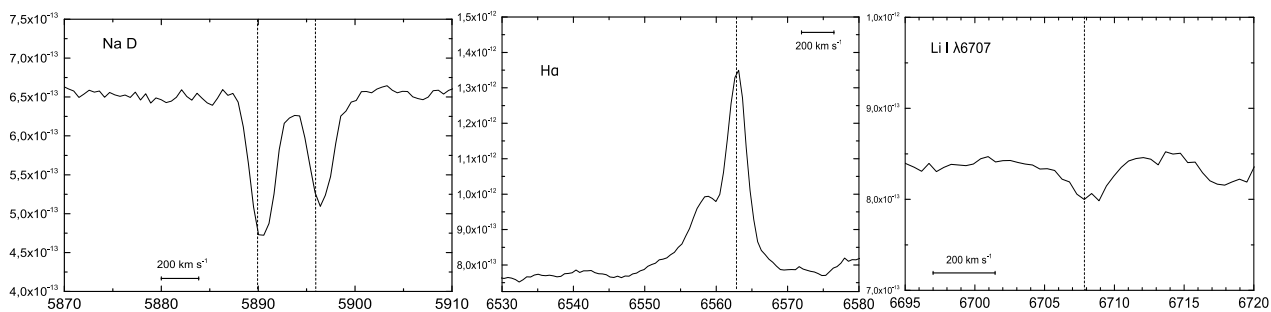}
   \caption{Enlarged fragments of the spectrum of 2MASS J05422123+223647, showing the profiles of Na D, H$\alpha$ and \ion{Li}{i} lines. Note the sharpness of the sodium lines and the split in lithium line. The laboratory wavelengths are indicated by short-dashed lines.}
   \label{spec}
\end{figure*}

 We measured equivalent widths and radial velocities for several lines to  compare them with the numerical data and the H$\alpha$ line profile, published by  \cite{Torres}.  Their similarity is remarkable. In fact, there have been only minimal changes over the past 30 years. We present several enlarged fragments in  Fig.\ref{spec}.

The structure of H$\alpha$ emission line remains double-peaked although the central depression no longer reaches continuum. The equivalent width of H$\alpha$  emission is 4~\AA\ (3~\AA\ in 1995). Its full width corresponds to $\approx 900\ $km s$^{-1}$  (800\ km s$^{-1}$ previously). The radial velocities (this paper uses heliocentric velocities, because the systemic radial velocity is small -- see later)  of H$\alpha$ components are $-$7\ km s$^{-1}$ (redward emission), $-$180\ km s$^{-1}$ (blueward emission) and $-$139\ km s$^{-1}$ (central depression). The corresponding values, roughly estimated from the Fig.2b of  \cite{Torres}, are $+50, -185$ and $-$81 km s$^{-1}$. One can also note the definite existence of one more emission component in the blue wing of H$\alpha$ with the radial velocity about $-$450\ km s$^{-1}$ \citep[it probably can be seen also in the spectrum shown by][]{Torres}.

The split of \ion{Li}{I} absorption is also clearly visible in our spectra. The equivalent width of \ion{Li}{I} $\lambda$ 6708 \AA\ absorption line is 0.25 \AA, and its components are split by about 45  km s$^{-1}$. The radial velocity of this line as whole is about +7\ km s$^{-1}$. This value should be compared with $-$6 km s$^{-1}$, measured by \cite{Torres}. It can be assumed that the systemic radial velocity is small and within measurement errors. On the other hand, NaD lines in the spectrum of 2MASS J05422123+223647 are narrow and sharp, with +20\ km s$^{-1}$ radial velocity, and do not show any traces of internal structure.
More detailed analysis of line profiles and radial velocities requires spectra of higher resolution.

\begin{figure*}[h!]
  \centering
  \includegraphics[width=0.8\textwidth]{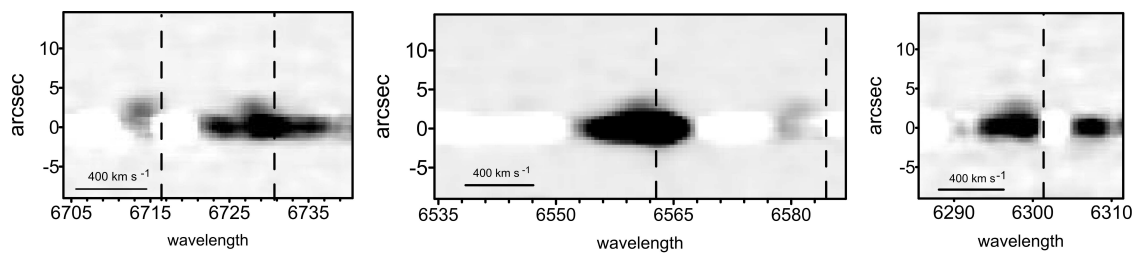}
   \caption{Emission lines of the jet from RNO 54 star:  [\ion{S}{ii}], H$\alpha$ + [\ion{N}{ii}] and [\ion{O}{i}]. The continuum spectrum of the star is subtracted. The broad H$\alpha$ emission in the stellar spectrum also is prominent; other details in continuum are only the fluctuations after subtraction.   The laboratory wavelengths are indicated by dashed lines. The  blue-shift of the jet emissions is clearly seen. }
   \label{jet}
\end{figure*}

An unexpected result of our long-slit spectroscopy was the discovery of the short and faint emission jet near the northern side of the star. The jet is visible  in the H$\alpha$, [\ion{N}{ii}], [\ion{S}{ii}] and [\ion{O}{i}]  lines, which confirms its  shock-excited nature. Most likely, the jet is oriented along the axis of the elliptic nebula. Its length does not exceed 4.5\arcsec. The intensity of jet emissions is so low that on Fig.\ref{jet}, where these emissions are shown, they are fainter even in comparison with the fluctuations in the subtracted stellar continuum.  No evidences of counterjet as well as of the inner structure of the jet emissions were found.

We measured the radial velocity of each emission line of the jet separately. All forbidden lines ([\ion{O}{i}] $\lambda$ 6300 \AA, [\ion{N}{ii}] $\lambda$ 6583 \AA, \ion{[S}{ii}] $\lambda\lambda$ 6716 and 6731 \AA) have virtually the same radial velocity with a mean value of $-119 \pm 3$ km
s$^{-1}$. The  H$\alpha$ emission in the jet has somewhat lower radial velocity:  $-85 \pm 18$ km
s$^{-1}$. The mean FWZI (full width at zero intensity) of the jet emission is $193 \pm 19$  km
s$^{-1}$ for the forbidden lines and 265 km s$^{-1}$ for H$\alpha.$ This difference in widths and velocities may be attributed to the well-known fact \citep[e.g.,][]{RB} that in the bow-shocks in the head of flows, where the   H$\alpha$ emission is brighter, it has the large velocity dispersion. Probably the mixing of jet emission and dust-scattered stellar  H$\alpha$ emission also may have effect on the radial velocity. Anyway, the velocity of jet is very close to the radial velocity of the central depression in H$\alpha$ line.  

\section{Discussion and conclusions}

The characteristics of 2MASS J05422123+223647 (H$\alpha$ emission, conspicuous \ion{Li}{I} line, shock-excited jet,   loop-like nebula indicating the presence of circumstellar dust disk) fully confirm its PMS nature. However, one needs to know its luminosity to define more precise classification.
One must also take into account that if this star indeed is a FU Ori type object after outburst, it cannot be considered as a "normal" star. In fact, according to the currently most successful model \citep[e.g.][and references therein]{HK,audard} it should be surrounded by a dense disk, which plays a role of extended atmosphere. The split of Li line can be attributed to existence of just such disk (see examples in \citealt{MHC,HCM}).

 We have compiled the photometric estimates in optical range, available from various sources, in the Table \ref{photometry}. One should note an excellent agreement between the photometry of \citet{Torres} and the APASS data as well as the obvious discrepancy of Tycho 2 catalogue with others: brightness in B and V differs for 1.5-2 mag and even B-V index is lower for 1 mag in Tycho 2. The reason of this discrepancy may be due to actual photometric variability of the star as well as to measurement errors (caused, for example, by close nebulosity around the star, which, very probably, affected, on the other hand, the photographic photometry in GSC-2.3 catalogue). This question should be solved by further photometric observations. 

The only long-term photometric observations of this object were performed  during the INTEGRAL-OMC programme \citep{OMC}. These observations show that the brightness of the star remains remarkably constant during the period from 2003 to 2010 with a mean value of V = 13.12 mag and  variations ranging from 12.94 to 13.25 mag. This value fully agrees with the APASS  and \citet{Torres} data. The recent search in OMC archive shows, that after the accumulating 5821 data points up to the 2022 the behaviour of the star did not change. Only 2 single measurements with V $\approx\ $12.4  and  11.4 stand out from the others and their reliability is questionable. Based on these data we conclude that 2MASS J05422123+223647 has very minimal photometric variability, if any at all.

To evaluate its absolute magnitude in optical range one must know the distance, approximate spectral type and interstellar reddening.
But the various estimates of  the interstellar extinction for this object give controversial results. The extinction map of \citet{FR} in this point gives $A_V \approx 1.6-2.0$ mag. The appearance of the stellar continuum also points to existence of the  moderate extinction on the line of sight. Assuming the spectral type of the star as G0 II and taking the intrinsic colours from the table of \citet{S-K}, we derived $A_V \approx 2.5-3.0$ mag from the BV photometry of \cite{Torres} and APASS. In comparison, photometry from Tycho 2 catalogue suggests a very low (near zero)
 value of $A_V$, which seems improbable.

The distance of 2MASS J05422123+223647 currently is determined fairly reliably  from Gaia EDR3 parallax. According to the catalog of \cite{BJ}, its distance estimate are 1387 pc (geometric) and  1396 pc (photogeometric). Using the trigonometric parallax value from Gaia DR3 (0.7163$\pm$0.0238 mas) and applying the systematic error correction, we obtain 1360$\pm$40 pc.  All these estimates well agree with each other, notwithstanding the slightly high value of 1.46 for RUWE (renormalised unit weight error), which probably is a result of the close nebulosity around the star.   

Assuming 1400 pc for the distance of the star and $A_V$ = 2.7 mag, we get $M_V = -0.3$ for $V = 13.12,$  or $L_V\approx110L_\sun $ correspondingly. Of course, these are only lower limit estimates, because this object demonstrates significant near and middle IR emission.  Besides, the true $A_V$ value may be higher, if consider the very possible circumstellar extinction.
 In any case, these values put the central star of RNO 54 between II and III luminosity classes, and its jet will be of linear size about 6000 AU.
   
The recent astrophysical data, presented in Gaia DR3 and computed by the modelling of the BP/RP spectra, suggest even higher luminosity of 2MASS J05422123+223647. In particular,   the following parameters should be noted: $T_{eff}=6700 K$, log $g=2.32$,  extinction $A_V = 4.14^m$ \cite[we obtained this value converting $A_{RP}$ according to][]{WangChen}, $M_G = -2.48$ and $L= 778L_\sun$. But these data are based on the assumption that modelled stars are ``normal'', which is definitely not the case for   2MASS J05422123+223647. Thus, the distance of 1913 pc, computed from this model, obviously is exaggerated, as well as absolute magnitude and luminosity of the star.

\begin{table}                                                            
  \caption{2MASS J05422123+223647 photometry}                                               \label{photometry}                                                       \centering                                                              \begin{tabular}{l c c c c c c c}                                                  \hline\hline
\cite{Torres} & U & B & V & R & I\\
\hline
& 15.35 & 14.75 &  13.12 & 12.01 & 10.91 \\
 \hline\hline
  Catalogues & B & V & g & r & i \\
  \hline                                                                   UCAC4\tablefootmark{a} & 14.73 & 13.22 &  & 12.45 & 11.55\\
  APASS DR10 & 14.71 & 13.05 & 13.93 &12.42 &11.56\\
  Tycho 2 & 12.05 & 11.44 & \\
  GSC-2.3\tablefootmark{b} & 12.23 & 11.51 & \\
  PS1 & & &\  13.79 & 13.34 & 12.04  \\
   \hline\hline
   Gaia & G & BP & RP & Plx \\
   \hline
   DR1 & 11.92 \\
   DR2 & 12.14 & 13.31 & 11.05 & 0.6496 \\
   EDR3 \& DR3\  & 12.12 & 13.29 & 11.04 & 0.7163 \\
  \hline
   \end{tabular} 
                         
\tablefoot{\\
\tablefoottext{a}{UCAC4  catalogue uses photometry data from APASS DR9.}\\
\tablefoottext{b}{The photometry in GSC-2.3 catalogue is based on photographic plates.}
}
 \end{table}

\begin{figure}[h!]
  \centering
  \includegraphics[width=0.5\textwidth]{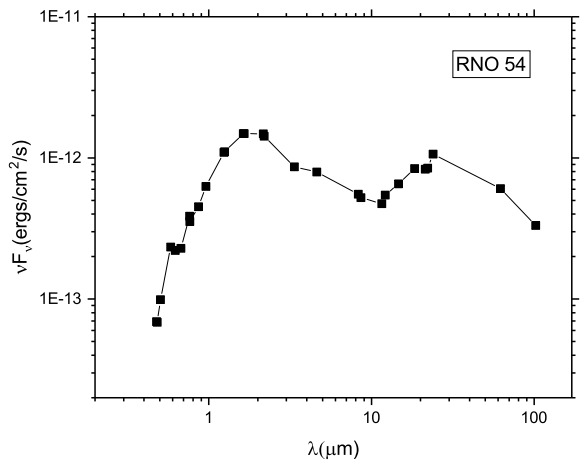}
   \caption{The SED of 2MASS J05422123+223647, built on the base of the data, taken from PanSTARRS, Gaia, SDSS, 2MASS, WISE, AKARI, MSX and IRAS surveys. }
   \label{SED}
\end{figure}

Using the Vizier Photometry viewer we constructed the SED of this object (Fig.\ref{SED}) and compared it with the SED, presented in the work of \cite{Torres}. The new SED is based on larger amount of data points, but has remarkable similarity to the previous one. As can be seen, the star shows obvious IR excess, which, however, drops in the far-IR wavelengths, making its SED typical for Class II sources, or, according to the models of \citet{RWI}, for middle-mass object in the evolutionary Stage II with significant contribution of disk emission. We estimated the lower limit of its bolometric luminosity by integrating the SED, and obtained $L_{bol}= 250L_\sun$ value. It should be noted, that \cite{Torres} also made the similar estimation from their SED, obtaining a value about $1.5-2L_\sun$ for an arbitrary distance of 100 pc. Taking into account the real distance of RNO 54, we derive the value of $L_{bol}= 300-400L_\sun$. On the base of all mentioned above we believe that $300 L_\sun$  can be reasonable conservative estimate for the bolometric luminosity of RNO 54 central object. Of course, it includes as the radiation from  the star itself as well as from the circumstellar disk. The value obtained is lower that mean  luminosity of HAeBe stars (the spectral type also is too late); on the other hand,  the star is much more luminous than T Tau stars.

The question naturally arises if this object indeed belongs to FUori-like objects, i.e. to FUors long after the outburst, as it was suggested by \cite{Goodrich} and \cite{Torres}. We consider this hypothesis as a very probable. The arguments for  this assumption are the following: spectral type (late F-early
G), the narrow absorptions and other spectral features (e.g. \ion{Ba}{ii} blend near $\lambda$6496 \AA), typical for supergiants and particularly for FUors, split of \ion{Li}{I} line, the absence of emission lines except of weak H$\alpha$, virtually no changes in the star brightness. The most important factors, in our opinion, are the existence of the shock-excited collimated outflow and the significant bolometric luminosity, which is characteristic for many FUor and FUor-like objects \cite[see, e.g.][]{audard}.
One
can see that 2MASS J05422123+223647  properties satisfy four from seven main FUor features, listed in \citet{Magakian} (but taking also into account that no outburst was ever observed).  

The only significant counterargument to such classification is the absence of conspicuous, wide and strongly blueshifted P Cyg type absorptions, which usually are observed in the spectra of FUors, but are not visible in the 2MASS J05422123+223647 spectrum.
However, recent studies of FUors, FUor-like stars and intermediate objects reveal the noticeable variety of their spectra. The small and rather faint jet observed in RNO 54 system, as well as the lack of more distant Herbig-Haro knots may indicate  that the star has been in a post-outburst state after the FUori event for a significant time (possibly several hundred years or more), during which the intensity of outflow  substantially lowered. Thus, it could have lost some of the characteristics typically associated with classical FUors.
Unfavorable orientation of the remains of expanding shell toward the line of sight cannot be excluded either.

No observational data yet are available to check if CO absorption bands exist in the near-IR spectrum of RNO 54 and if the gradual change in its spectral type with wavelength can be detected. The existence of its low-amplitude variability also should be analyzed with the aid of more precise photometry. We also have no explanation for the out-of-line data of Tycho 2 mission.

To summarise, we tend to think that the discovery of the faint emission jet and confirmation of significant luminosity to a great extent
 increase the likelihood that RNO 54 star belongs to FUor-like objects.

\begin{acknowledgements}
            Authors are grateful to referee for many helpful comments, which improved the paper. Authors thank Prof. A.S. Rastorguev for  valuable advice. This work was partly supported by RA
State Committee of Science, in the frames of the research project 21T-1C031. 

The spectroscopic observations on the unique scientific facility ``Big Telescope Alt-azimuthal'' of SAO RAS, as well as the data reduction, were financially supported by grant No 075-15-2022-262 (13.MNPMU.21.0003) of the Ministry of Science and Higher Education of the Russian Federation.

This work uses results from the European Space Agency (ESA) space mission Gaia. Gaia data are being processed by the Gaia Data Processing and Analysis Consortium (DPAC). Funding for the DPAC is provided by national institutions, in particular the institutions participating in the Gaia MultiLateral Agreement (MLA). This work also uses the data from the OMC Archive at CAB (INTA-CSIC), pre-processed by ISDC and further processed by the OMC Team at CAB.
The OMC Archive is part of the Spanish Virtual Observatory project. Both are funded by MCIN/AEI/10.13039/501100011033 through grants PID2020-112949GB-I00 and PID2019-107061GB-C61, respectively. This research has made use of the VizieR catalogue access tool, CDS,
 Strasbourg, France.  

\end{acknowledgements}

\end{document}